\documentclass[aps,prl,twocolumn,groupedaddress]{revtex4}

\usepackage{graphicx}% Include figure files
\usepackage{dcolumn}% Align table columns on decimal point
\usepackage{bm}% bold math

\begin{document}

%Title of paper
\title{Dispersion of "Dispersionless Band":\\
Comments on L. Brey and H.A. Fertig paper Electronic States of Graphene Nanoribbons Studied With The Dirac Equation}

\author{Lyuba Malysheva}
\author{Alexander Onipko}
\email[]{aleon@ifm.liu.se}
\thanks{This work was partly supported by the Special Program of the Section of 
Physics and Astronomy of NANU and the Visby program (SI).}
\affiliation{Bogolyubov Institute for Theoretical Physics, 03680 Kyiv, Ukraine}

\date{\today}

\begin{abstract}
As a particular application of the earlier proposed model of graphene as a macromolecule, we have found the exact analytical expression of dispersion relation for the band of edge states in graphene zigzag ribbons. This band is often referred to as "dispersionless band" or "zero modes". The obtained results contrast conclusions regarding edge states given in the referenced paper and show where the earlier given description is valid and where it is not correct. 
\end{abstract}

% insert suggested PACS numbers in braces on next line
%\pacs{73.22.-f}

\maketitle

\section{Introduction}\label{Sec1}

Dramatic effects of zigzag edges on the band structure and density of states of finite-size layers of 2D graphite were demonstrated by first-principles calculations in 1993 \cite{Kobayashi}. In many subsequent studies, peculiar graphene properties associated with edge states, have been studied in many details by different methods \cite{Fujita,Nakada,W,Ez,Peres2,Peres1,Brey,Sa,prb2007}. In particular, exponential behavior of the density of states and dispersion near the Fermi energy has attracted much of attention. An exponential decrease of edge state energy observed in numerical studies \cite{Kobayashi,Nakada}, was explained analytically by Brey and Fertig \cite{Brey}, henceforth referenced as BF. In this paper, electron states of graphene ribbons (GRs) were studied with the use of the Dirac equation, that is within a relativistic model \cite{Ando}, originating from the tight binding description of infinite 2D graphite \cite{Wallace,Saito,Thomsen}. To take into account edge effects, certain boundary conditions have been imposed on the wave function, separately, for A and B sublattices indicated in Fig.~1a. An impressive quantitative agreement between tight-binding calculations and solutions of the Dirac equations was demonstrated by a number of illustrative examples. However, from the methodological point of view, such an approach suffers from certain weaknesses, that will be addressed here later on.

As mentioned, the edge-state spectrum  of zigzag GRs was quantified by BF in an analytical form. It relates the decrease of energy of edge states with an increase of separation of longitudinal wave vector $k^{\rm BF}_y$ from a zero-energy point of 2D graphite dispersion relation. One of these so-called K points is shown in Fig.~1. According to BF, the spectrum of edge states is described by two approximate equations, reproduced below by Eqs.~(\ref{b1}) and (\ref{b2}) in slightly different notations. 

Here we show that this result is correct only partly. In fact, for the larger part of edge states, the edge-state energy is governed by an exponential behavior with different pace and crossover from one exponential dependence for smaller wave vectors to another for larger wave vectors. These dependencies, represented in Eqs.~(\ref{b4}) and (\ref{b11}), follow from the exact treatment of graphene as a macromolecule \cite{Lyuba}. Our equation (\ref{b4}) agrees but Eq.~(\ref{b11}) disagrees with the earlier prediction. We see the reason of the divergence in that the BF approximate description is not good enough for zigzag edges and therefore it does not catch all peculiarities of electronic properties in a very narrow region near the Fermi energy, where the edge states occur. 

\section{How Does Electron Energy Go to Zero in Zigzag Graphene Ribbons?}

To begin with, let us write the well known linear dispersion relation for an infinite graphene sheet $\pm\frac{\sqrt{3}}{2}ta\sqrt{(k^{\rm BF}_x)^2+ (k^{\rm BF}_y)^2}$, where $t$ is the hopping integral between nearest-neighbor carbons in the honeycomb lattice. Label BF indicates that the reference point in {k}-space (K point of the 2D graphite band structure) is the same as in the referenced paper. According to BF, for a graphite sheet of finite width $L=\sqrt{3}a(N-1/3)$ and an infinite length (a zigzag GR as it appears in Fig.~1a) the dispersion relation changes. Namely, if longitudinal component $k^{\rm BF}_y$ exceeds the critical value $k_y^c=1/L$, the transverse wave-vector component becomes imaginary, $ k^{\rm BF}_x =ik$,
\begin{equation}\label{b1}
\varepsilon = \pm\frac{\sqrt{3}}{2}ta\sqrt{ (k^{\rm BF}_y)^2-k^2},\quad k^{\rm BF}_y>k_y^c,
\end{equation}
and satisfies equation
\begin{equation}\label{b2}
k\coth(kL) = k^{\rm BF}_y.
\end{equation}
This is a slight modification of Eq.~(6) as it appears in Ref. \cite{Brey}. 

It is easy to verify that under the condition $k^{\rm BF}_y>2k_y^c$, the dependence of real solution to Eq.~(\ref{b2}) on $k ^{\rm BF}_y$ is accurately (with error less than 1\%) reproduced by
\begin{equation}\label{b3}
k=k ^{\rm BF}_y(1-2e^{-2k ^{\rm BF}_yL}).
\end{equation}
The use of this result in Eq.~(\ref{b1}) yields
\begin{equation}\label{b4}
\varepsilon=\pm \sqrt{3}t ak^{\rm BF}_y\exp(-k^{\rm BF}_yL)=\pm \sqrt{3}tq\exp(- \sqrt{3}Nq),
\end{equation}
which finalizes a fully analytical description of the edge-state spectrum. Here and henceforth, $N$ is assumed to be a large number in all approximate relations. The second equality represents a new notation, $ak^{\rm BF}_y\equiv q$, which is more suitable for further consideration. A point to note is that the divergence between Eq.~(\ref{b4}) and energies calculated from Eqs.~(\ref{b1}) and (\ref{b2}) is noticeable only for $k^{\rm BF}_yL<2$, see inset in Fig.~2. Common limitations of these two approximations will be clarified via comparison with the exact results discussed next.

\begin{figure}[htbp]
\includegraphics[width=0.45\textwidth]{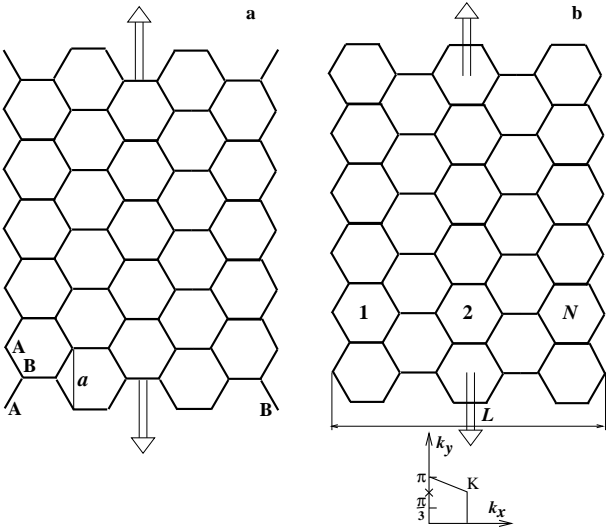}
\caption{Zigzag graphene ribbon of width $L=\sqrt{3}a(N-1/3)$ and infinite length, represented by (a) section of honeycomb lattice, periodic in A and B sites (2D graphite model), and (b) by $N$ C-C coupled polyacene chains (macromolecule model); $a$ is the minimal translation distance. Cross on $k_y$ axis indicates zero-energy point of graphene dispersion relation (\ref{b5}). Counted from this point, wave vector $(k_x,q/a)$ is equivalent to $(k^{\rm BF}_x,k^{\rm BF}_y)$. Quarter of the hexagonal Brillouin zone of 2D graphite is also shown.}
\end{figure}

The edge-state spectrum can be treated with the use of the model shown in Fig.~1b. In the framework of canonical tight-binding description, the band structure of zigzag GRs is determined by \cite{Lyuba} 
\begin{eqnarray}\label{b5}
&&E^\pm =\pm t\nonumber\\
&&\times\sqrt{1\pm4\left|\cos({ak_y}/{2})\cos ( \sqrt{3}ak_x/2)\right|+4\cos^2({ak_y}/{2})},\nonumber\\
\end{eqnarray}
where the upperscript $\pm$ refers to the sign plus or minus under the root, $k_y$ has the meaning of the longitudinal wave vector, $0\le ak_y\le\pi$, and values of $ak_x\equiv \kappa/\sqrt{3}$ should be found from equation
\begin{equation}\label{b6}
\frac{\sin \kappa^\pm N}{\sin \kappa^\pm (N+1/2)}=\mp2\cos(ak_y/2).
\end{equation}
Note that near zero-energy point, ${\bf k}=\frac{2\pi}{a}(0,\frac{1}{3})$,  the minus branch of Eq.~(\ref{b5}) can be rewritten as
\begin{equation}\label{b7}
E^- = \pm\frac{\sqrt{3}}{2}t\sqrt{(q=ak_y-2\pi/3)^2+(\kappa^-/3)^2}, 
\end{equation}
making obvious the correspondence between BF's and our notations: $|ak^{\rm BF}_x|=|ak_x|=\kappa^-/\sqrt{3}$, and $ak^{\rm BF}_y=q=ak_y-2\pi/3$. 

For the minus branch of dispersion relation (\ref{b5}), Eq.~(\ref{b6}) supports only real solutions $\kappa^-_\nu$, $\nu$=\,0,1,2,...,$N$$-$1,  if 0$\le$ $ak_y$$\le$$2\pi/3$+$q^c$, $q^c$=\,2arccos$[N/(2N$+1)]$-$2$\pi/3$. However, if $q$$>$$q^c$, Eq.~(\ref{b6}) has $N$$-$1 real and one imaginary solution. The latter is denoted below as $\kappa^-_0$. For $N$$>>$1, the critical value of the wave vector, $q^c$=$(\sqrt{3}N)^{-1}$$<<$1, coincides with $ak_y^c$ in the BF theory. 

It is the smallest of $k_y$-dependent solutions $\kappa^-_\nu$ that becomes imaginary, $\kappa^-_0(k_y)\rightarrow i\delta(q)$. In this case, electron energies are bound to the interval $|E|/t\le (2N)^{-1}$, where 
\begin{equation}\label{b8}
E^-=\pm t\frac{\sinh(\delta/2)}{\sinh (N+1/2)\delta}, 
\end{equation}
and $\delta$ satisfies
\begin{equation}\label{b9}
\frac{\sinh N\delta }{\sinh(N+1/2) \delta}= \cos(q/2)-\sqrt{3}\sin(q/2). 
\end{equation}
Henceforth, this energy interval is referred to as the band of edge states. Distinct from Eqs.~(\ref{b1}) and (\ref{b2}), Eqs.~(\ref{b8}) and (\ref{b9}) describing this band are exact. The dispersion of edge states predicted by the exact and approximate equations is represented in Fig.~2. 

For $q^c\le q<<1$, the band of edge states is reproduced by Eqs.~(\ref{b1})--(\ref{b2}) reasonably well. Equation (\ref{b4}) gives nearly equally good description excluding the interval $1\le q/q^c\le 2$. However, within the larger part of the actual interval $q^c\le q\le\pi/3$, the divergence between the approximate and exact descriptions is dramatic. Here, we mean the difference between functional forms $\varepsilon(q)$ and $E^-(q)$ rather than the divergence in numbers.

To specify an analytical expression of edge-state energy for larger values of $q$, one can solve Eq.~(\ref{b9}) in the same approximation that leads to Eq.~(\ref{b4}). For $N\delta>>1$, the solution to this equation can be expressed as $\delta =-2\ln\left[\cos(q/2)-\sqrt{3}\sin(q/2) \right]$. In fact, this approximate expression gives fairly accurate results for $N\delta>2$ and can be rewritten in a more compact form 
\begin{equation}\label{b10}
\delta =-2\ln\left[2\sin[(\pi/3-q)/2] \right],
\end{equation}
showing that when the wave vector approaches its maximal value, $q\rightarrow \pi/3$, the electron (hole) energy goes to zero as
\begin{equation}\label{b11}
E^- =\pm t(\pi/3-q)^{2N}.
\end{equation}
In contrast, according to Eqs.~(\ref{b1})--(\ref{b2}), the edge-state energy has though small but finite energy at $q=\pi/3$. In general, the obtained dependence $\delta(q)=\sqrt{3}k(ak^{\rm BF}_y$=$q)$ is pronouncedly different from its analogue in the BF description. Nevertheless, Eq.~(\ref{b4}) can be retrieved from Eqs.~(\ref{b8}) and (\ref{b10}) under the condition $q^c<< q<<1$, where the restriction on $q$ from above sets the limit of applicability of Eq.~(6) in Ref.~\cite{Brey}.  

Despite rather crude approximations made under the passage from Eqs.~(\ref{b8}) and (\ref{b9})  to Eq.~(\ref{b11}), the latter works reasonably well even for $q\approx0.3$, see Fig.~2. The comparison of the exact, Eq.~(\ref{b8}), and approximate, Eq.~(\ref{b4}) and Eq.~(\ref{b11}), dependencies in this figure illustrates two regions of the wave vector, where electron/hole dispersion is qualitatively different, although it is exponential in both regions. The crossover from one exponential behavior to another occurs within an interval that can be approximately designated as $\pi/12<q<\pi/6$. This specific feature of edge-state spectrum is totally lost in the BF description, illustrated by the curve (4) in Fig~2.

\begin{figure}[htbp]
\includegraphics[width=0.45\textwidth]{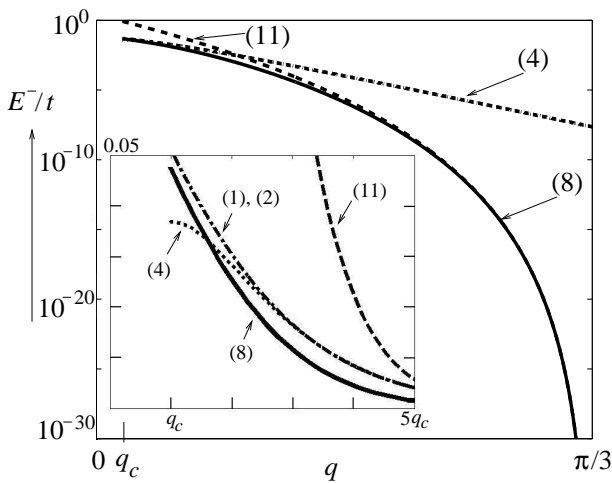}
\caption{Energy of electron edge states in a $N=10$ wide zigzag nanoribbon (semi-logarithmic scale). Exact and approximate dependencies on the longitudinal wave vector are marked by corresponding equations. Curves calculated from (\ref{b1}), (\ref{b2}) and, alternatively, from (\ref{b4}) are undistinguishable in this scale. Inset: same data for small values of $q$ in non-logarithmic scale. }
\end{figure}

In connection with the above discussion, frequently used terms such as "zero modes", "dispersionless modes", and "partly flat bands" can be quantified. As shown, there is only one electron/hole state with zero energy at $k_y=\pi/a$. It is characteristic for infinitely long graphene sheets with zigzag edges and belongs to the lowest/highest electron/hole band of in total $2N$ electron/hole dispersion branches. The spectrum of any finite stripe of zigzag GR contains no zero-energy levels at all. For example, in a 10 nm wide and 100 nm long zigzag graphene nanoribbon, the lowest/highest electron/hole level is $E^-_{min/max}\approx t\,10^{-97}$ (in this estimate, $a=1.42\sqrt{3}$ \AA \cite{Saito}). The term "dispersionless mode or band" refers, in fact, to an exponential dispersion within an interval $t(2N)^{-1}\approx t/[4.7L({\rm nm})+2/3]$, up and down the Fermi level. The exponential dependence of edge-state energy on the wave vector changes from $\sim \exp(-\sqrt{3}Nq)$ to $\sim \exp(2N\ln(\pi/3-q)$ within an interval specified above. The precise quantification of the "flat part" of the lowest/highest conduction/valence band is given by Eqs.~(\ref{b8}) and (\ref{b9}).

\section{2D Graphite or Macromolecule Model?}
The above presented results refer to two models of graphene which are totally equivalent, as an object of Physics \cite{Note}, but require different formal treatments. The 2D graphite model of graphene (a) (see \cite{Ando,Saito,Thomsen}) has been exploited by BF. This is a periodic structure of two nonequivalent atoms A and B. The alternative model (b) is equivalent to a 2D crystal with four nonequivalent atoms. Therefore, the corresponding dispersion relations (or band structures) are different \cite{Lyuba}. As long as the bulk properties of infinite graphene are in focus, both models and both descriptions are equivalent. But once boundaries and/or size effects come into play, the graphene properties are more easily and naturally described starting from the macromolecule model rather than 2D graphite model. This is demonstrated by the present consideration and by obtaining accurate analytical expressions of the band structure for achiral graphene ribbons and carbon nanotubes \cite{Lyuba3}. The use of the Dirac equations demands some boundary conditions which can be formulated on the grounds of general requirements (such as current conservation, etc.)  \cite{Mc,Been}. However, from our point of view, there is a controversy in any attempt to rationalize boundary effects with the use of imposed boundary conditions. The use of the macromolecule model is free of this and other difficulties of approximate descriptions.

As already mentioned the possibility to use the Dirac equation for a description of graphene electronic properties comes out from 2D graphite model of graphene. This means that the dispersion relation is obtained with the use of periodic boundary conditions for A and B sublattices. A commonly spread belief is that the dispersion relation is {\it independent} of the boundary conditions. But the case of graphene presents an exception. Modeled as a real macromolecule, the graphene dispersion relation has only two zero-point energies (instead of six at the corners of the hexagonal Brillouin zone) neither of which has {\bf k} coordinates coinciding with the hexagon corners \cite{Lyuba}. Therefore, references to K points in discussions of graphene electronics should be used with precaution. 

In summary, it is shown that for the larger part of the band of edge states in zigzag graphene ribbons, the description given on the basis of the k-p approximation and corresponding Dirac equations agrees poorly with the basic tight-binding model. In particular, such description does not identify $ak_y = \pm \pi$ as the only points, where the conduction and valence bands touch each other, and it does not reproduce correctly the exponential decrease (increase) of the edge-state electron (hole) energy to zero value. We have suggested an exact analytical description, based on the same model assumptions. For the larger part of the band in focus this alternative description is expressed in elementary functions. As a continuation of this work, accurate analytical expressions for low- (high-) lying electron (hole) bands have been obtained for zigzag and armchair graphene ribbons and carbon nanotubes \cite{Lyuba3}. The use of the k-p approximation for these structures becomes thus unnecessary.


\begin{thebibliography}{99}
\bibitem{Kobayashi}
K. Kobayashi, Phys. Rev. B {\bf 48}, 1757 (1993).

\bibitem{Fujita}
M. Fujita, K. Wakabayashi, K. Nakada, and K. Kusakabe,
J. Phys. Soc. Japan {\bf 65}, 1920 (1996). 

\bibitem{Nakada}
K. Nakada, M. Fujita, G. Dresselhaus, and M.~S. Dresselhaus,
Phys. Rev. B {\bf 54}, 17954 (1996).

\bibitem{W} 
K. Wakabayashi, M. Fujita, H. Ajiki, and M. Sigrist,
Phys. Rev. B {\bf 59}, 8271 (1999).

\bibitem{Ez}
M. Ezawa, Phys. Rev. B {\bf 73}, 045432 (2006). 

\bibitem{Peres2}
N.~M.~R. Peres, F. Guinea, and A.H. Castro Neto, 
 Phys. Rev. B {\bf 73} 125411 (2006). 

\bibitem{Peres1}
N.~M.~R. Peres, A.~H. Castro Neto, and F. Guinea,
 Phys. Rev. B {\bf 73}, 195411 (2006). 

\bibitem{Brey}
L. Brey and H.~A. Fertig, Phys. Rev. B {\bf 73}, 235411 (2006). 

\bibitem{Sa}
K.-I. Sasaki, S. Murakami, and R. Saito,
 J. Phys. Soc. Japan {\bf 75} 074713 (1996). 

\bibitem{prb2007} 
M. Kohmoto and Y. Hasegava, Phys. Rev. B {\bf 76} 205402 (2007). 

\bibitem{Ando}
T. Ando,
J. Phys. Soc. Japan {\bf 74}, 777 (2002). 

\bibitem{Wallace}
P.~R. Wallace,  Phys. Rev. B {\bf 9}, 622 (1947).

\bibitem{Saito}
R.Saito, G. Dresselhaus, and M.~S. Dresselhaus, {\it Physical Properties of Carbon Nanotubes} (imperial College, London, 1998).

\bibitem{Thomsen}
S. Reich, C. Thomsen, and J. Maultzsch {\it Carbon Nanotubes: Basic concepts and Physical Properties} (WILEY-VCH, Weinheim, 2004).

\bibitem{Lyuba}
L.~Malysheva and A.~Onipko, Phys. Rev. Lett., 100, 186806 (2008); arXiv:0801.4155v1 [cond-mat.mes-hall].
\bibitem{Note}
Obviously, two models shown in Fig.~1 are equivalent in the limit ${\cal N}\rightarrow\infty$. However, for a stripe of finite, 2D graphite model (a) can be considered as graphene macromolecule with defects and {\it vice versa}, graphene macromplecule model (b) is just defected 2D graphite; that is (a) and (b) are different systems.

\bibitem{Lyuba3}
L. Malysheva and A. Onipko, arXiv:0803.1761v1 [cond-mat.mes-hall].

\bibitem{Mc}
E. McCann and V.I. Fal'ko, J. Phys. Condens. Matter {\bf 16}, 2371 (2004).

\bibitem{Been}
A. R. Akhmerov, C. W. J. Beenakker, arXiv/0710.2723v3 [cond-mat.mes-hall].




\end{thebibliography}
\end{document}